\documentclass[aps,twocolumn,prd,showpacs,nofootinbib]{revtex4}

\usepackage{graphicx}
\usepackage{amssymb}
\usepackage{amsmath}
\usepackage[mathscr]{eucal}
\input{colordvi.tex}

\newcommand{\sss}[1]{{\scriptscriptstyle{#1}}}

\newcommand{\uPl}{\mathrm{Pl}}

\newcommand{\usssPl}{\sss{\uPl}}

\newcommand{\ie}{\textrm{i.e.}~}

\newcommand{\mpl}{m_\usssPl}

\newcommand{\bear}{\begin{array}}  \newcommand{\eear}{\end{array}}
\newcommand{\bea}{\begin{eqnarray}}  \newcommand{\eea}{\end{eqnarray}}
\newcommand{\beq}{\begin{equation}}  \newcommand{\eeq}{\end{equation}}
\newcommand{\bef}{\begin{figure}}  \newcommand{\eef}{\end{figure}}
\newcommand{\bec}{\begin{center}}  \newcommand{\eec}{\end{center}}

\newcommand{\del}{\partial}

\begin{document}

\title{DBI-essence}

\author{J\'er\^ome Martin} \email{jmartin@iap.fr} \affiliation{ Institut
d'Astrophysique de Paris, UMR 7095-CNRS, Universit\'e Pierre et Marie
Curie, 98bis boulevard Arago, 75014 Paris, France}

\author{Masahide Yamaguchi} \email{gucci@phys.aoyama.ac.jp}
\affiliation{Department of Physics and Mathematics, Aoyama Gakuin
University, Sagamihara 229-8558, Japan}

\date{\today}

\begin{abstract}
Models where the dark energy is a scalar field with a non-standard
Dirac-Born-Infeld (DBI) kinetic term are investigated. Scaling solutions
are studied and proven to be attractors. The corresponding shape of the
brane tension and of the potential is also determined and found to be,
as in the standard case, either exponentials or power-law of the DBI
field. In these scenarios, in contrast to the standard situation, the
vacuum expectation value of the field at small redshifts can be small in
comparison to the Planck mass which could be an advantage from the model
building point of view. This situation arises when the present-day value
of the Lorentz factor is large, this property being {\it per se}
interesting. Serious shortcomings are also present such as the fact
that, for simple potentials, the equation of state appears to be too far
from the observational favored value $-1$. Another problem is that,
although simple stringy-inspired models precisely lead to the power-law
shape that has been shown to possess a tracking behavior, the power
index turns out to have the wrong sign. Possible solutions to these
issues are discussed.
\end{abstract}

\maketitle

\section{Introduction}
\label{sec:intro}

Since the discovery that the expansion of the Universe is presently
accelerated~\cite{Perlmutter:1998np,Riess:1998cb,Tegmark:2003ud,
Spergel:2006hy,Fosalba:2003ge,Scranton:2003in,Boughn:2003yz,Solevi:2004tk,
Mainini:2005mq,Fuzfa:2006ps}, various suggestions have been made in
order to explain this observational fact. Amongst them is the hypothesis
of dark energy, a fluid with a negative pressure representing about
$70\%$ of the total energy density in the Universe. The question of the
physical nature of the dark energy has, of course, been widely
discussed. The most natural candidate, still perfectly compatible with
all the data available, is the cosmological constant. However, the
difficulty to reconcile the value of $\Lambda$ deduced from the
observations with the value calculated
theoretically~\cite{Weinberg:1988cp} (maybe too naively?) has prompted
the study of alternatives. Clearly, a simple scalar field, a
``quintessence'' field, is a natural candidate for such an
alternative~\cite{Ratra:1987rm,Ferreira:1997hj,Brax:1999gp,
Brax:1999yv,Brax:2000yb}. Amongst all the possibilities, scalar fields
with inverse power-law potentials have attracted lot of interest
because, in this case, there is a solution of the equations of motion
that is an attractor~\cite{Ratra:1987rm}. This means that the
present-day behavior of the Universe is insensitive to the initial
conditions. Usually, the attractor solution is a scaling solution, \ie a
solution for which the energy density scales as a power of the scale
factor~\cite{Liddle:1998xm}.

\par

If the above mentioned route is correct, then another interesting issue
is whether a candidate for quintessence in high energy physics can be
identified. Clearly, this cannot be done without going beyond the
standard model of particle physics. In particular, it would be very
interesting to achieve this goal in string theory since it is presently
our best candidate as a unified theory~\cite{Brax:2007me}.

\par

Recently, there have been many works aiming at connecting string theory
with inflation which is also a phase of accelerated expansion (but
taking place in the very early Universe at a much higher energy
scale). For this purpose, new ideas in string theory based on the
concept of branes have revealed themselves especially fruitful. In
particular, scenarios where the inflaton is interpreted as the distance
between two branes moving in the extra dimensions along a warped throat
have given rise to many interesting
studies~\cite{Kachru:2003sx,Silverstein:2003hf,Alishahiha:2004eh,
Lorenz:2007ze}. In this article, we want to investigate whether the same
kind of ideas can lead to sensible dark energy scenarios.

\par

At the technical level, scenarios of the type mentioned above lead to
scalar field models where the kinetic term is non-canonical. More
precisely, the kinetic term has a Dirac-Born-Infeld (DBI)
form. Physically, this originates from the fact that the action of the
system is proportional to the volume traced out by the brane during its
motion. This volume is given by the square-root of the induced metric
which automatically leads to a DBI kinetic term. Therefore, as a first
step toward a scenario of ``DBI-essence'', it is first necessary to
understand whether scaling and attractor solutions are still present
when the scalar field has a DBI kinetic term. This question constitutes
the main target of the present article.

\par

This paper is organized as follows. In Sec.~\ref{sec:scaling}, we
briefly review the scaling properties of a quintessence field with a
standard kinetic term. Then, in Sec.~\ref{sec:scalingdbi}, we reconsider
this question but with a DBI kinetic term. In particular, we compare the
DBI results with the standard ones. In Sec.~\ref{sec:numerics}, we study
the behavior of DBI-essence at small redshifts. Since, in this case,
the scalar field is no longer a test field, this requires numerical
computations. Finally, in Sec.~\ref{sec:conclusions}, we present our
conclusions and discuss the open issues that should be studied in the
future.

\section{Scaling Solutions with a Standard Kinetic Term}
\label{sec:scaling}

We consider a spatially flat Friedmann-Lemaitre-Robertson-Walker (FLRW)
Universe containing a perfect fluid and a scalar field $\phi$. Assuming
that the scalar field and the perfect fluid are separately conserved,
the equations of motion are given by
\begin{eqnarray}
\label{eq:friedmann}
H^2 &=& \frac{\kappa }{3}\left(\rho+\rho _{\phi}\right)\, ,
\\
\label{eq:conservfluid}
\dot{\rho}+3H\left(\rho +p\right)&=& 0\, ,
\\
\label{eq:conservscalar}
\dot{\rho }_{\phi}+3H\left(\rho _{\phi}+p_{\phi}\right)&=& 0\, ,
\end{eqnarray}
where $H\equiv \dot{a}/a$ is the Hubble parameter and $\rho$ and $p$ are
respectively the energy density and the pressure of the perfect fluid. A
dot denotes a derivative with respect to cosmic time and the quantity
$\kappa $ is defined by $\kappa \equiv 8\pi /\mpl^2$. In the following,
we assume that the perfect fluid has a constant equation of state
parameter $\omega \equiv p/\rho$, the two cases of main interest being
$\omega =1/3$ for the radiation-dominated era and $\omega =0$ for the
matter-dominated era. In this case, the conservation
equation~(\ref{eq:conservfluid}) can be integrated exactly and leads to
the familiar behavior $\rho \propto a^{-3(1+\omega )}$. Moreover, if we
further assume that the scalar field is a test field and that the
evolution of the background geometry is mainly controlled by the perfect
fluid, then one has $a(t)\propto t^{2/[3(1+\omega )]}$ or, for the
Hubble parameter, $H=2/\left[3(1+\omega)t\right]$.

\par

Let us first briefly remind what are the scaling solutions in the simple
case where the scalar field has a standard kinetic term. In this
situation, the energy density and the pressure are given by the familiar
expressions
\begin{equation}
\label{eq:rhopstandard}
\rho _{\phi}=\frac{\dot{\phi }^2}{2}+V(\phi )\, ,
\quad p_{\phi}=\frac{\dot{\phi }^2}{2}
-V(\phi)\, .
\end{equation}
If one inserts these expressions into the conservation equation for the
scalar field~(\ref{eq:conservscalar}), then one obtains the Klein-Gordon
equation
\begin{equation}
\label{eq:kgstandard}
\ddot{\phi}+3H\dot{\phi }+V'(\phi)=0\, ,
\end{equation}
where a prime denotes a derivative with respect to $\phi $. Then, we
seek potentials $V(\phi )$ such that the energy density of the test
scalar field scales as a power-law of the scale factor, namely $\rho
_{\phi}\propto a^{-3(1+\omega _{\phi})}$ where $\omega _{\phi}\equiv
p_{\phi}/\rho _{\phi}$ is a constant. It has been established in
Ref.~\cite{Ratra:1987rm} that scaling solutions exist if the potential
has an exponential shape,
\begin{equation}
V(\phi)=M^4{\rm e}^{-\lambda \phi}\, ,
\end{equation}
where $\lambda $ is a constant or is of the Ratra-Peebles type (\ie
inverse power-law of the field), namely
\begin{equation}
\label{eq:potrp} 
V(\phi)=M^{4+\alpha }\phi^{-\alpha }\, .
\end{equation}
In the first case, the particular solution leading to the scaling
behavior reads
\begin{equation}
\label{eq:attraexpstandard}
\phi(t)=\frac{2}{\lambda }\ln \left(\frac{t}{t_0}\right) \, ,
\end{equation}
the constant $\lambda $ and the mass scale $M$ being linked by the
relation
\begin{equation}
\label{eq:standardMexp}
\lambda ^2M^4t_0^2=\frac{2(1-\omega )}{1+\omega }\, .
\end{equation}
As is well-known, the quintessence equation of state parameter is just
given by the equation of state of the background perfect fluid, $\omega
_{\phi}=w$. The particular solution~(\ref{eq:attraexpstandard}) is
important because it is an attractor. This means that the final (\ie
present day) evolution of the field is in fact independent of the
initial conditions. At the technical level, this can be seen by studying
small (linear) perturbations around the attractor. The eigenvalues of
the perturbations around the critical point can be expressed as
\begin{equation}
\label{eq:lambdaexp}
\lambda _{\pm}=\frac{1}{2m}\left[\left(m-6\right)\pm 
\sqrt{\left(m-6\right)^2+8m\left(m-6\right)}\right]\, ,
\end{equation}
where we have defined $m\equiv 3(1+\omega)$. Since $\omega <1$, one has
$m<6$ and the eigenvalues are negative and one has a stable spiral
point. Moreover, for the particular
solution~(\ref{eq:attraexpstandard}), one has
\begin{equation}
\label{eq:standardattraexp}
\frac{{\rm d}^2V}{{\rm d}\phi^2}=\frac{9}{2}
\left(1-\omega _{\phi}^2\right)H^2\, .
\end{equation}
This is an important formula because it implies that $\phi \sim \mpl $
today. Indeed, $V''\sim V/\phi ^2$ and $H^2\sim V/\mpl ^2$ when the
field starts dominating the energy density content of the Universe;
equating these two quantities leads to the above mentioned
conclusion. For this reason, a sensible model building of quintessence
is only possible in a supergravity (SUGRA)
framework~\cite{Brax:1999gp,Brax:1999yv}. However, a well-known
difficulty of the exponential case is that the property $\omega
_{\phi}=\omega $ implies that the scalar field cannot drive an
accelerated expansion. This is why the inverse power-law case seems to
be more interesting.

\par

In the case of the Ratra-Peebles potential~(\ref{eq:potrp}), there also
exists an exact particular solution of the Klein-Gordon equation that is
an attractor. It reads
\begin{equation}
\label{eq:particularrp}
\phi=\phi_0\left(\frac{t}{t_0}\right)^{2/(\alpha +2)}\, ,
\end{equation}
where the quantity $\phi_0$ is linked to the mass scale $M$ by the formula
\begin{equation}
\label{eq:standardMrp}
M^{4+\alpha }\phi_0^{-\alpha -2}t_0^2=\frac{2}{\alpha (\alpha +2)}\left(
\frac{2}{1+\omega }-\frac{\alpha }{\alpha +2}\right)\, .
\end{equation}
For this particular solution, the equation of state parameter can be
expressed as
\begin{equation}
\label{eq:eosrp}
\omega _\phi=\frac{\alpha \omega -2}{\alpha +2}\, .
\end{equation}
As expected, in the limit $\alpha \rightarrow +\infty$, one recovers the
exponential case, $\omega _{\phi}=\omega$. However, the crucial
difference with the exponential potential is that one can now have
$\omega _{\phi}<\omega$, that is to say, the scalar field energy density
can now scale more slowly than the background fluid and, hence,
eventually dominates, causing the Universe to accelerate. Moreover, the
solution~(\ref{eq:particularrp}) is also an attractor as revealed by a
dynamical system analysis. Indeed, the eigenvalues of small
perturbations around the critical point read
\begin{equation}
\label{eq:lambdarp}
\lambda _{\pm}=\frac{\left(2n-m-6\right)\pm 
\sqrt{\left(2n-m-6\right)^2+8m\left(n-6\right)}}{2m}\, ,
\end{equation}
where we have defined $n\equiv 3(1+\omega_{\phi})$. Again, one can show
that there exists a stable spiral point as long as both eigenvalues are
negative, which is equivalent to $2n-m-6 < 0$. On this attractor, the
evolution of the second order derivative of the potential is given by
\begin{equation}
\label{eq:standardattrarp}
\frac{{\rm d}^2V}{{\rm d}\phi^2}=\frac{9}{2}
\frac{\alpha +1}{\alpha }\left(1-\omega _{\phi}^2\right)H^2\, ,
\end{equation}
and one check that this last equation reproduces the corresponding
equation in the exponential case when $\alpha \rightarrow
+\infty$. Again, this prompts a SUGRA treatment of the model building
issue since one still has $\phi \sim \mpl $ now.

\section{Scaling Solutions with a DBI Kinetic Term}
\label{sec:scalingdbi}

Let us now consider that the dark energy scalar field is a
Dirac-Born-Infeld (DBI) scalar field. In this case, the action of the
field can be written as
\begin{equation}
\label{eq:actiondbi}
S_{_{\mathrm{DBI}}}=-\int {\rm d}^4x a^3(t)\left[T(\phi)
\sqrt{1-\frac{\dot{\phi}^2}{T(\phi)}}+V(\phi)-T(\phi)\right]\, ,
\end{equation}
where $T(\phi)$ is the tension and $V(\phi)$ is the potential. From this
expression, it is easy to obtain the corresponding energy density and
pressure of the scalar field. They read
\begin{equation}
\label{eq:rhopdbi}
\rho _{\phi}=\left(\gamma -1\right)T(\phi)+V(\phi)\, ,
\quad
p_{\phi}=\frac{\gamma -1}{\gamma}T(\phi)-V(\phi)\, ,
\end{equation}
where the quantity $\gamma $ is reminiscent from the usual relativistic
Lorentz factor and is given by
\begin{equation} 
\gamma \equiv \frac{1}{\sqrt{1-\dot{\phi}^2/T(\phi)}}\, .
\end{equation}
The expressions~(\ref{eq:rhopdbi}) of the energy density and pressure of
the DBI field should be compared to their standard counterpart, see
Eqs.~(\ref{eq:rhopstandard}). As usual, if one inserts
Eqs.~(\ref{eq:rhopdbi}) in the conservation
equation~(\ref{eq:conservscalar}), one obtains the DBI Klein-Gordon
equation, namely
\begin{equation}
\label{eq:kgdbi}
\ddot{\phi}-\frac{3T'(\phi)}{2T(\phi)}\dot{\phi}^2+T'(\phi)
+\frac{3H}{\gamma ^2}\dot{\phi}+\frac{1}{\gamma ^3}
\left[V'(\phi )-T'(\phi)\right]=0\, .
\end{equation} 
We notice that the equation of motion for $\phi$ is quite complicated
compared to Eq.~(\ref{eq:kgstandard}) despite the fact that the
conservation equation has retained its standard form.

\par

Let us also compare with other works in the literature. Let us start
with K-essence where the action can be written as
\begin{equation}
S=\int {\rm d}^4x \sqrt{-g}\, p\left(\phi ,X\right)\, ,
\end{equation} 
where $X=\left(\nabla \phi\right)^2/2$. Clearly the
action~(\ref{eq:actiondbi}) is a special case of the above
action. However, as first discussed in Ref.~\cite{Chiba:1999ka},
K-essence usually means that the potential term vanishes and the
negative pressure of the scalar field is realized only by considering
the kinetic term~\cite{Garriga:1999vw,ArmendarizPicon:2000dh,
ArmendarizPicon:2000ah,Chiba:2002mw}. On the other hand, our model
cannot realize the negative pressure without the potential term, as
shown below. Therefore, this class of models cannot encompass
Eq.~(\ref{eq:actiondbi}). Another model related to the present study is
the case where the dark energy field is a tachyon for which the action
is given by~\cite{Gorini:2003wa}
\begin{equation}
S=-\int {\rm d}^4x \sqrt{-g}\,V(T)\sqrt{1+\frac{1}{M^4}g^{\mu \nu}
\partial _{\mu }T\partial _{\nu}T}\, ,
\end{equation} 
where $M$ is a fundamental scale and $V(T)$ is a potential which, of
course, needs not to be the same function as $V(\phi)$ in
Eq.~(\ref{eq:actiondbi}). This class of theory is equivalent to the case
studied here (through a redefinition of the field) only when the
potential in Eq.~(\ref{eq:actiondbi}) vanishes. Let us also notice that
when $V(T)$ is constant the model is in fact equivalent to the Chaplygin
gas with the equation of state $p\propto
-1/\rho$~\cite{Fabris:2001tm,Gorini:2002kf}. Therefore, beside the fact
that the search of scaling solutions has not yet been investigated in
this type of models, we conclude that the class of scenarios under
scrutiny in this paper was not considered before.

\par

As a warm up, let us find the scaling solutions in the simple case where
the potential vanishes. As already mentioned before, this means that we
now seek tensions $T(\phi )$ such that the energy density of the test
DBI scalar field scales as a $\rho _{\phi}\propto a^{-3(1+\omega
_{\phi})}$. From Eqs.~(\ref{eq:rhopdbi}), it is easy to show that
$\gamma $ is, in this case, constant and given by $\gamma =1/\omega
_{\phi}$. Then, the formula of the energy density, $\rho _{\phi}=
(\gamma -1)T(\phi)\propto a^{-3(1+\omega _{\phi})}$ immediately gives
the scaling in time of $T(\phi)$ which in turn, combined with $\dot{\phi
}^2/T(\phi )=(\gamma ^2-1)/\gamma ^2$ and the fact that $\gamma $ is
constant, implies that
\begin{equation}
\label{eq:particularsoldbi}
\dot{\phi }\propto t^{-(1+\omega _{\phi})/(1+\omega)}\, ,
\end{equation}
This equation is easily solved. Let us start with $\omega _{\phi}=\omega
$. In this case, one has $\phi \propto \ln t$ and, as a consequence,
\begin{equation}
\label{eq:tensionexp}
T(\phi )=M^{4}{\rm e}^{-\lambda \phi }\, ,
\end{equation}
where $\lambda $ is a constant. Again, this case is very similar to the
situation where we have a standard kinetic term and a exponential
potential. As a consequence, this model suffers from the standard
phenomenological problems. Since the scalar fields exactly tracks the
background matter, one cannot have a large enough contribution of dark
energy density today without spoiling Big Bang Nucleosynthesis
(BBN). Moreover, the scalar field behaves as matter today and,
therefore, cannot cause the acceleration of the Universe.

\par

On the other hand, if $\omega _{\phi }\neq \omega $, then the scalar
field is just a power-law of the cosmic time which implies that $T(\phi
)$ can be expressed as
\begin{equation}
\label{eq:tensionrp}
T(\phi )=M^{4+\alpha}\phi ^{-\alpha}\, ,
\end{equation}
where $M$ is a mass scale and $\omega _{\phi}$ is related to $\alpha $
and the background equation of state parameter $\omega $ through the
relation
\begin{equation}
\omega _{\phi }=\frac{\alpha \omega -2}{\alpha +2}
=\frac{1}{\gamma }\, .
\end{equation}
Interestingly enough, as discussed before, this is exactly the equation
obtained when there is a standard kinetic term with a Ratra-Peebles
potential, see Eq.~(\ref{eq:eosrp}). However, in the present case,
$\omega _{\phi}=1/\gamma >0$. This means that the solution is physically
relevant only if $\omega >2/\alpha$ which excludes the case $\omega =0$,
at least for $\alpha >0$.

\par

Since it appears that the previously described situation is not
satisfactory, we now envisage the case where the potential $V(\phi)$ is
non vanishing. In order to deal with this problem, we assume that
$\gamma $ is a constant. Without this hypothesis, the problem is
technically very complicated but the actual convincing argument in favor
of this assumption is that the corresponding scaling solutions (with
$\gamma $ constant) are attractors, see below. Then, the crucial
observation is that Eq.~(\ref{eq:particularsoldbi}) is still valid
because, in its derivation, one has never assumed that $V=0$. This
implies that, as in the case of a vanishing potential, scaling solutions
exist for tensions $T(\phi)$ given by Eq.~(\ref{eq:tensionexp}) or
Eq.~(\ref{eq:tensionrp}). Then, the Klein-Gordon
equation~(\ref{eq:kgdbi}) can be used to determine the
potential. Straightforward manipulations lead to
\begin{equation}
\label{eq:ratioVT}
\frac{V(\phi)}{T(\phi)}=\frac{\gamma ^2-1}{\gamma }
\left(\frac{1}{1+\omega _{\phi}}-\frac{\gamma}{1+\gamma}\right)\, ,
\end{equation}
that is to say the potential is proportional to the tension and has also
the exponential shape or inverse power-law shape. It is interesting to
notice that, when the field is on tracks in the standard kinetic case,
the potential term is also proportional to the kinetic term, that is to
say the ratio of the potential term to the kinetic term $K \equiv
\dot{\phi}^2/2$ is a constant given by $V(\phi)/K(\phi) =
2/(1+\omega_{\phi})-1$.

\par

In the case of an exponential potential, the exact solution (from now
on, we put a subscript ``e'' to denote the quantities that are evaluated
with the exact particular solution of the Klein-Gordon equation) reads
$\phi _{\rm e}(t)=2/\lambda \ln(t/t_0)$ with $\lambda ^2M^4t_0^2=4\gamma
_{\rm e}^2/(\gamma _{\rm e}^2-1)$ and the equation of state parameter is
$\omega _{\phi}=\omega$. In the Ratra-Peebles case, one as $\phi _{\rm
e}(t)=\phi _0(t/t_0)^{2/(\alpha +2)}$ with
\begin{equation}
M^{4+\alpha }\phi _0^{-\alpha -2}t_0^2=\frac{4\gamma _{\rm e}^2}{
(\alpha +2)^2(\gamma _{\rm e}^2-1)}\, ,
\end{equation}
and the equation of state has the standard form given by
Eq.~(\ref{eq:eosrp}). It is importance to notice that, because we deal
with a modified Klein-Gordon equation, the expressions of $\lambda
^2M^4t_0^2$ and $M^{4+\alpha }\phi _0^{-\alpha -2}t_0^2$ are different
from the ones obtained previously, see Eqs.~(\ref{eq:standardMexp})
and~(\ref{eq:standardMrp}). Let us also remark that these formulas can
either be obtained from the requirement that the Lorentz factor is
constant or by brute force calculation using the Klein-Gordon equation.
 
\par

Let us now study the behavior of small perturbations around the
particular solutions. Let us first start with the exponential case. For
this purpose, we rewrite the equation of motion in terms of $u(\tau )$
defined by $u\equiv \lambda(\phi-\phi _{\rm e})$ and $t\equiv {\rm
e}^{\tau}$. If we write $p\equiv u'$, then one obtains the system
\begin{widetext}
\begin{eqnarray}
\frac{{\rm d}p}{{\rm d}\tau}&=&-\left(5+\frac{2}{\gamma
^2}\frac{1}{1+\omega}\right)p-\frac{3}{2}p^2-\frac{4}{1+\omega}
\left(\frac{1}{\gamma^2}-\frac{1}{\gamma _{\rm e}^2}\right)
+\lambda ^2M^4\left({\rm e}^{-u}-1\right)
+\lambda ^2M^4\left[\frac{\gamma _{\rm e}^2-1}{\gamma _{\rm e}}
\left(\frac{1}{1+\omega }-\frac{\gamma _{\rm e}}{1+\gamma _{\rm e}}\right)
-1\right]\nonumber \\ & & \times 
\left(\frac{{\rm e}^{-u}}{\gamma ^3}-\frac{1}{\gamma _{\rm
e}^3}\right)=0\, ,\\
\frac{{\rm d}u}{{\rm d}\tau}&=&p \, ,
\end{eqnarray}
\end{widetext}
where, now, the quantity $\gamma $ is no longer a constant and can be
written as
\begin{equation}
\gamma =\left[1-\frac{\gamma _{\rm e}^2-1}{\gamma _{\rm e}}{\rm e}^u
\left(1+\frac{p}{2}\right)^2\right]^{-1/2}\, .
\end{equation}
As a consequence, we see that the critical point is
$(p,u)=(0,0)$. Notice that, for the critical point, one checks that
$\gamma =\gamma _{\rm e}$. We now consider the behavior of small
perturbations $(\delta p, \delta u)$ around the critical point
$(0,0)$. It is straightforward to establish that
\begin{eqnarray}
\label{eq:matrixexp}
\frac{{\rm d}}{{\rm d}\tau }
\begin{pmatrix}
\displaystyle
\delta p \cr \delta u
\end{pmatrix}
&=&
\begin{pmatrix}
1-\frac{2}{1+\omega } & 2-\frac{2}{1+\omega }
\frac{1+\gamma _{\rm e}^2}{\gamma _{\rm e}^2} \cr
1 & 0
\end{pmatrix}
 \begin{pmatrix}
\delta p \cr \delta u
\end{pmatrix}\, .
\end{eqnarray}
Then, the eigenvalues of this matrix can be expressed as
\begin{eqnarray}
\lambda _{\pm}&=&\frac{1}{2m}\Biggl[\left(m-6\right)
\nonumber \\
& & \pm\sqrt{\left(m-6\right)^2+8m\left(m-3\frac{\gamma _{\rm e}^2+1}{\gamma
_{\rm e}^2}\right)}\Biggr]\, .
\end{eqnarray}
This expression should be compared with Eq.~(\ref{eq:lambdaexp}). The
only difference is the presence of the factor $\gamma _{\rm e}$ in the
last term inside the square root. Otherwise, and this is quite
remarkable, the expression is the same. Let us also notice that the
condition that the kinetic energy cannot exceed the total energy is
equivalent to $n \le 3(\gamma_{\rm e}+1)/\gamma_{\rm e}$. Since
$\gamma_{\rm e} \ge 1$, the condition $n \le 3(\gamma_{\rm
e}^2+1)/\gamma_{\rm e}^2$ is stronger than the condition $n \le
3(\gamma_{\rm e}+1)/\gamma_{\rm e}$ and, therefore, is not automatically
satisfied in our case. We conclude that there is a stable spiral point
if $n \le 3(\gamma_{\rm e}^2+1)/\gamma_{\rm e}^2$.

\par

Let us now turn to the inverse power-law case. This time, the
dimensionless function $u(\tau )$ is defined by $u(\tau )\equiv
\phi/\phi _{\rm e}$, the definition of the time $\tau $ remaining the
same. Then, a straightforward calculation leads to the following system
of equations
\begin{widetext}
\begin{eqnarray}
\frac{{\rm d}p}{{\rm d}\tau}&=&-\left(\frac{5\alpha +2}{\alpha +2}
+\frac{2}{\gamma
^2}\frac{1}{1+\omega}\right)p-\frac{3\alpha}{2}\frac{p^2}{u}
-\frac{4}{1+\omega}\frac{u}{\alpha +2}
\left(\frac{1}{\gamma^2}-\frac{1}{\gamma _{\rm e}^2}\right)
+\alpha M^{4+\alpha}\phi _0^{-\alpha -2}\left(u^{-\alpha -1}-u\right)
\nonumber \\ & &
+\alpha M^{4+\alpha}\phi _0^{-\alpha -2}
\left[\frac{\gamma _{\rm e}^2-1}{\gamma _{\rm e}}
\left(\frac{1}{1+\omega }-\frac{\gamma _{\rm e}}{1+\gamma _{\rm e}}\right)
-1\right]
\left(\frac{u^{-\alpha -1}}{\gamma ^3}-\frac{u}{\gamma _{\rm
e}^3}\right)=0\, ,\\
\frac{{\rm d}u}{{\rm d}\tau}&=&p \, ,
\end{eqnarray}
\end{widetext}
where, this time, the Lorentz factor can be written as
\begin{equation}
\gamma =\left\{1-\frac{u^{\alpha}}{M^{4+\alpha }\phi_0^{-\alpha -2}}
\left[p^2+\frac{4up}{\alpha +2}+\frac{4u^2}{\left(\alpha +2\right)^2}\right]
\right\}^{-1/2}\, .
\end{equation}
It is clear form the above system that the critical point is now given
by $(p,u)=(0,1)$. The next step is to study the behavior of small
perturbations $(\delta p,1+\delta u)$ around the critical point. One
arrives at
\begin{eqnarray}
\frac{{\rm d}}{{\rm d}\tau }
\begin{pmatrix}
\displaystyle
\delta p \cr \delta u
\end{pmatrix}
&=&
\begin{pmatrix}
\frac{\alpha -2}{\alpha +2}-\frac{2}{1+\omega } & \frac{2\alpha}{\alpha
 +2} -\frac{2}{1+\omega }\frac{1+\gamma _{\rm e}^2}{\gamma _{\rm e}^2} \cr
1 & 0
\end{pmatrix}
 \begin{pmatrix}
\delta p \cr \delta u
\end{pmatrix}\, .\nonumber \\
\end{eqnarray}
As expected, in the limit $\alpha \rightarrow +\infty$, the above matrix
exactly reproduces the matrix obtained in the exponential case, see
Eq.~(\ref{eq:matrixexp}). Then, the next step is to determine the
eigenvalues. The result reads
\begin{eqnarray}
\lambda _{\pm}&=&\frac{2n-m-6}{2m}\nonumber \\
&\pm&\frac{1}{2m}\sqrt{\left(2n-m-6\right)^2
+8m\left(n-3\frac{\gamma _{\rm e}^2+1}{\gamma
_{\rm e}^2}\right)}\, .
\end{eqnarray}
This expression should be compared with Eq.~(\ref{eq:lambdarp}). As it
was the case before, the modification introduced by the DBI kinetic term
is only apparent in the last term inside the square root. Therefore,
there is a stable spiral point if $n \le 3(\gamma_{\rm
e}^2+1)/\gamma_{\rm e}^2$ and $2n-m-6<0$ are satisfied.

\par

Finally, on the attractor, in the exponential case, it is easy to
establish that the following relation holds
\begin{equation}
\frac{{\rm d}^2V}{{\rm d}\phi^2}=9\gamma _{\rm e}
\left(1+\omega \right)\left[1-\left(1+\omega \right)
\frac{\gamma _{\rm e}}{1+\gamma _{\rm e}}\right]H^2 \, .
\label{eq:mass2expDBI}
\end{equation}
This formula is the generalization of
Eq.~(\ref{eq:standardattraexp}). Obviously, one can also establish the
corresponding expression in the inverse power-law case. It reads
\begin{equation}
\frac{{\rm d}^2V}{{\rm d}\phi^2}=9\frac{\alpha +1}{\alpha}\gamma _{\rm e}
\left(1+\omega _{\phi}\right)\left[1-\left(1+\omega _{\phi}\right)
\frac{\gamma _{\rm e}}{1+\gamma _{\rm e}}\right]H^2 \, ,
\label{eq:mass2powDBI}
\end{equation}
and this is equivalent to Eq.~(\ref{eq:standardattrarp}). This has
important consequences for model building. Indeed, if one repeats the
discussion after Eq.~(\ref{eq:standardattraexp}), then one arrives at
the conclusion that 
\begin{equation}
\phi \sim \frac{\mpl}{\sqrt{\gamma _{\rm e}}} 
\end{equation}
because the second term in the bracket in Eq. (\ref{eq:mass2powDBI})
cannot exceed unity. Therefore, if $\gamma _{\rm e}\gg 1$, then the
vacuum expectation value of the field is not necessarily large in Planck
units. This is certainly an important advantage of the DBI models over
the standard ones as respect to model building issues.

\par

Finally, it is also worth commenting about the shape of the tension
$T(\phi)$. From a stringy point of view, the inverse of $T(\phi)$
represents the warp factor of the throat in which the branes are
living. A natural choice~\cite{Alishahiha:2004eh} is $T(\phi)\propto
\phi ^4$, that is to say $\alpha =-4$. Therefore, this case belongs to
the class of tracking models considered here which is a non trivial
result. Unfortunately, the sign of the exponent is not the correct
one. Indeed, for $\alpha =-4$, one has $\omega _{\phi}=2\omega +1>\omega
$ which means that, despite the presence of an attractor, the scalar
field scales faster than the background fluid and, hence, can never
dominate the matter content of the Universe.

\section{Numerical Calculations}
\label{sec:numerics}

In this section, we investigate the behavior of the DBI scalar field at
small redshifts, when it starts dominating the matter content of the
universe. In this situation, the assumption that it is a test field
breaks down and numerical calculations are required. 

\begin{figure}[t]
\includegraphics[width=8cm, height=7cm, angle=0]{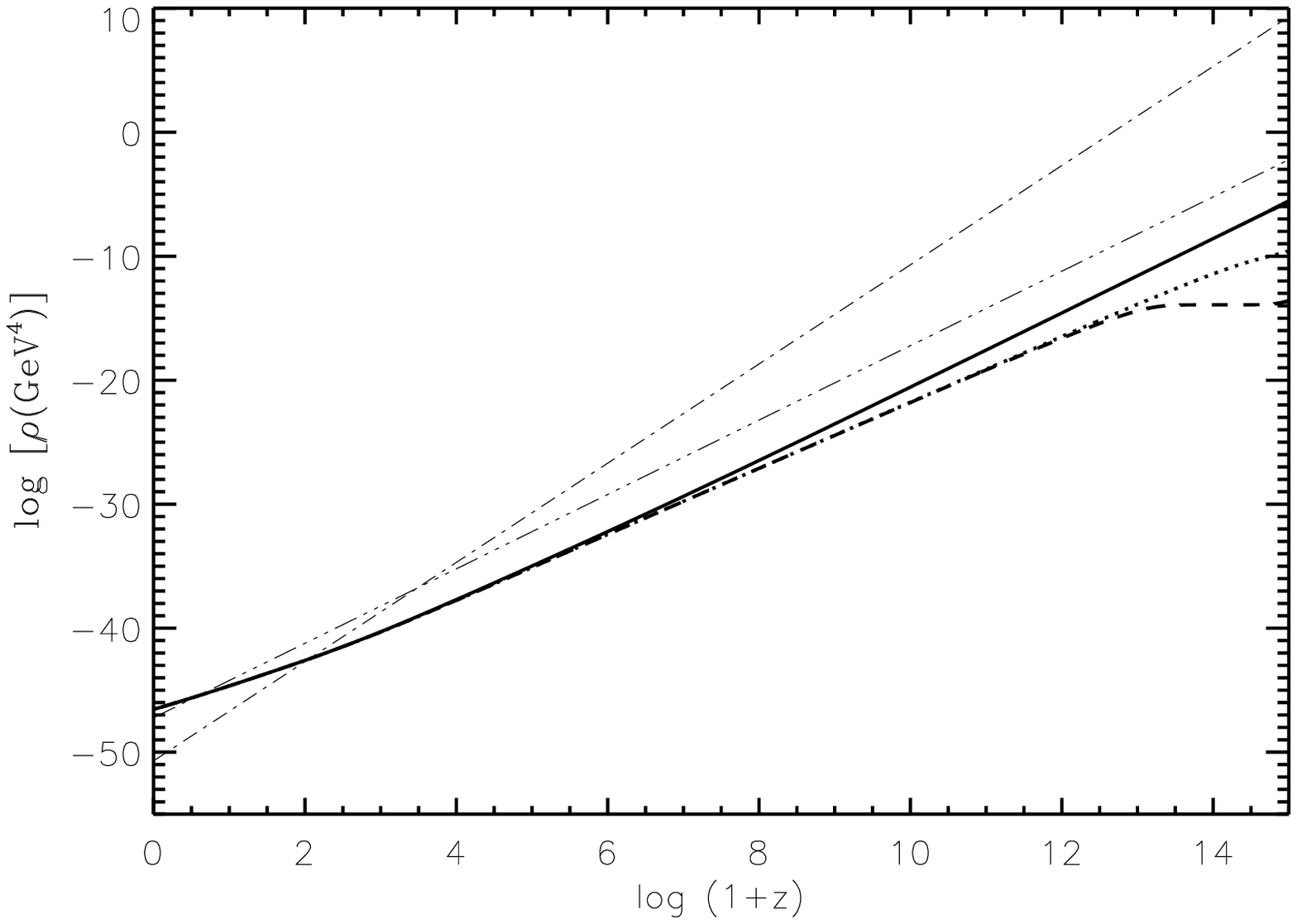}\\
\caption{Evolution of the DBI energy density for different initial
conditions in the case where the potential is of the Ratra-Peebles type
with $\alpha =4$. The value of $C$ is chosen to be $C\sim 3.443$ which
corresponds to $\gamma _{\rm e}=20$, see Eq.~(\ref{eq:defC}). The
initial velocity $\dot{\phi }_{\rm ini}$ is always chosen such that
$\gamma _{\rm ini}=5$. The solid line corresponds to an initial vacuum
expectation value of $\phi_{\rm ini}/\mpl\sim 10^{-10}$, the dotted line
to $\phi_{\rm ini}/\mpl\sim 10^{-9}$ and the dashed line to $\phi_{\rm
ini}/\mpl\sim 10^{-8}$. The energy density of radiation (dotted-dashed
line) and cold dark matter (dotted-dotted-dashed line) are also
represented.}  \label{fig:nrj}
\end{figure}

\begin{figure}[t]
\includegraphics[width=8cm, height=7cm, angle=0]{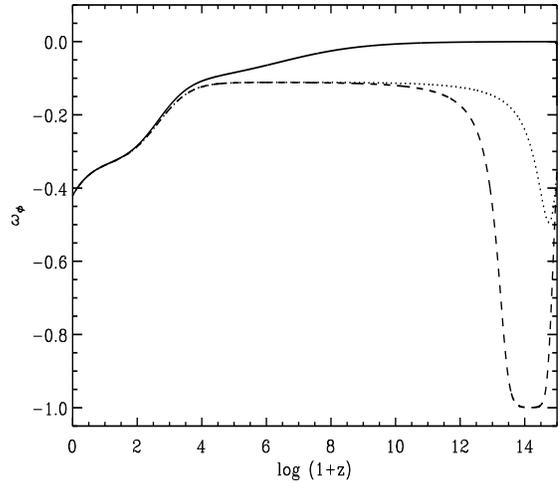}\\
\caption{Evolution of the DBI equation of state for different initial
conditions in the case where the potential is of the Ratra-Peebles
potential with $\alpha =4$. As in Fig.~\ref{fig:nrj}, the value of $C$
is chosen to be $C\sim 3.443$ which corresponds to $\gamma _{\rm e}=20$
and the initial velocity $\dot{\phi }_{\rm ini}$ is such that $\gamma
_{\rm ini}=5$. The solid line corresponds to an initial vacuum
expectation value of $\phi_{\rm ini}/\mpl\sim 10^{-10}$, the dotted line
to $\phi_{\rm ini}/\mpl\sim 10^{-9}$ and the dashed line to $\phi_{\rm
ini}/\mpl\sim 10^{-8}$. The final (present day) value of $\gamma $ is
$\gamma _0\sim 3.96$ and $\phi _0/\mpl\sim 1.37$. Finally, the equation
of state is such that $\omega _0\sim -0.42$, $\omega _1\sim 6.68\times
10^{-2}$.}  \label{fig:state}
\end{figure}

We first check that the attractor is observed numerically. As a
representative example, we have chosen to investigate the case $\alpha
=4$. In Fig.~\ref{fig:nrj}, we have represented the evolution of the DBI
energy density for three different initial conditions (more precisely,
the initial velocity is always the same and corresponds to an initial
value of the Lorentz factor $\gamma _{\rm ini}=5$ but different initial
vacuum expectation values $\phi _{\rm ini}$ are considered). In order to
have a DBI energy density today equal to $70\%$ of the critical energy
density, we have tuned the scale $M$ of the brane tension $T(\phi)$, see
Eq.~(\ref{eq:tensionrp}). The mass scale of the potential is determined
by Eq.~(\ref{eq:ratioVT}) which implies that $V(\phi)=C M^{4+\alpha}\phi
^{-\alpha}$ where $C$ is defined by
\begin{eqnarray}
\label{eq:defC}
C&\equiv &\frac{\gamma ^2_{\rm e}-1}{\gamma _{\rm e}}
\left(\frac{1}{1+\omega _{\phi}}-\frac{\gamma_{\rm e}}{1+\gamma_{\rm e}}
\right) \nonumber \\ &=&
\frac{\gamma ^2_{\rm e}-1}{\gamma _{\rm e}}
\left[\frac{\alpha +2}{\alpha(1+\omega)}
-\frac{\gamma_{\rm e}}{1+\gamma_{\rm e}}
\right]
\, .
\end{eqnarray}
Choosing a value of $C$ is in fact equivalent to choosing the value of
the Lorentz factor on the attractor, $\gamma _{\rm e}$, during a phase
of evolution characterized by the background equation of state $\omega
$. So, for instance, in Fig.~\ref{fig:nrj}, we have chosen $\gamma _{\rm
e}=20$ and $\omega =1/3$. This means that the attractor solution should
be such that $\gamma _{\rm e}=20$ during the radiation dominated
era. Given Eq.~(\ref{eq:defC}) and $\alpha =4$, this choice implies that
$C\sim 3.443$. In this case, there is also an attractor during the
matter dominated era but the corresponding value of the Lorentz factor
is different. It is easy to show that it reads
\begin{equation}
\label{eq:gammatter}
\gamma _{\rm e}^{\rm cdm}=\frac{\alpha (C-1)}{4}
\left[1+\sqrt{1+\frac{8(\alpha +2)}{\alpha ^2(C-1)^2}}\right]\, .
\end{equation}
In the present case, this gives $\gamma _{\rm e}^{\rm cdm}\sim 5.438$.

\par

The attractor behavior is clearly seen in Fig.~\ref{fig:nrj}. For
initial conditions $\phi_{\rm ini}/\mpl\sim 10^{-9}$ and $\phi_{\rm
ini}/\mpl\sim 10^{-8}$, the attractor is joined during the
radiation-dominated era while for $\phi_{\rm ini}/\mpl\sim 10^{-10}$, it
is reached during the matter-dominated era.

\par

In Fig.~\ref{fig:state}, we have represented the evolution of the
equation of state for the same situation. Again, the attractor behavior
is clearly noticed. We can even check numerically that, on the
attractor, Eq.~(\ref{eq:eosrp}) is valid. Since we consider a model with
$\alpha =4$, the DBI equation of state during the radiation dominated
era should be $\omega _{\phi}\simeq -0.11$. Clearly, this is what is
obtained in Fig.~\ref{fig:state}. The present day value of the equation
of state is $\omega _0\simeq -0.42$ (the derivative of the equation of
state at vanishing redshift being $\omega _1\simeq 6.68\times
10^{-2}$). The corresponding value for a scalar field with a standard
kinetic term and the same Ratra-Peebles potential is $\omega _0\simeq
-0.487$. Firstly, and contrary to a naive expectation, the equation of
state is not pushed towards $-1$. Therefore, it seems that we do not
gain anything in comparison with the model with a standard kinetic
term. Secondly, the value obtained seems to be too large given the
constraints available on $\omega _0$. Even if one should put a damper on
these constraints since they have not been obtained for the model under
considerations here (usually, a simple law of the form
$\omega=\omega_0+\omega _1z$ are used, which is clearly not valid for
the model under consideration here, and this can cause a ``bias
problem'', see Ref.~\cite{Virey:2004pr}), the value is so far from $-1$
that the model is probably in trouble from the observational point of
view.  This is clearly a very serious problem for the class of models
studied in the present article. One possibility is to decrease the value
of $\alpha$. For instance, $\alpha =0.3$ implies $\omega _{\phi}\sim
-0.9$. Of course, the corresponding model seems contrived and, in
addition, in this case, a small value of the equation of state would
also be obtained with a standard kinetic term. Another possibility would
be to consider other shapes for the tension and the potential. The new
shape of $T(\phi)$ and $V(\phi)$ should approximatively reduces to the
inverse power-law shape at large redshifts such that the attractor
behavior is preserved and should differ from it at small redshifts in
order to obtain an equation of state closer to $-1$. A typical example
is provided by the SUGRA
potential~\cite{Brax:1999gp,Brax:1999yv}. However, we show below that
even this solution could not work in the DBI case.

\begin{figure}[t]
\includegraphics[width=8cm, height=7cm, angle=0]{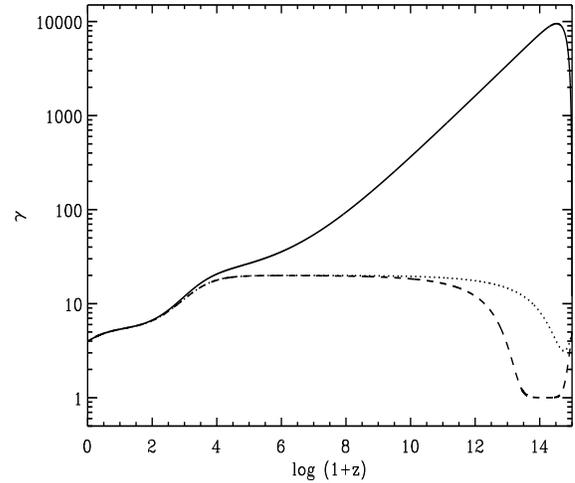}\\
\caption{Evolution of the Lorentz factor for different initial
conditions for the Ratra-Peebles potential with $\alpha =4$. The value
of $C$ is chosen to be $C\sim 3.443$ which corresponds to $\gamma _{\rm
e}=20$ and the initial velocity $\dot{\phi }_{\rm ini}$ is chosen such
that $\gamma _{\rm ini}=5$. The solid line corresponds to an initial
vacuum expectation value of $\phi_{\rm ini}/\mpl\sim 10^{-10}$, the
dotted line to $\phi_{\rm ini}/\mpl\sim 10^{-9}$ and the dashed line to
$\phi_{\rm ini}/\mpl\sim 10^{-8}$. The final value of the Lorentz factor
is $\gamma _0\sim 3.96$} \label{fig:gamma}
\end{figure}

\begin{figure}[t]
\includegraphics[width=8cm, height=7cm, angle=0]{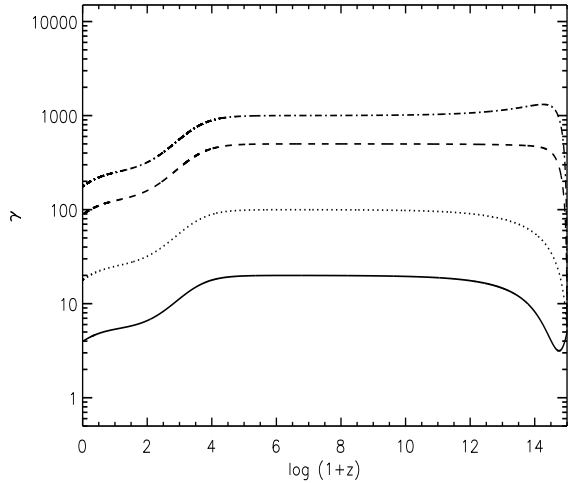}\\
\caption{Evolution of the Lorentz factor for different values of $\gamma
_{\rm e}$ in the case where the potential is of the Ratra-Peebles type
with $\alpha =4$. The values considered are $\gamma _{\rm e}=20$ (solid
line), $\gamma _{\rm e}=100$ (dotted line), $\gamma _{\rm e}=500 $
(dashed line) and $\gamma _{\rm e}=1000 $ (dotted-dashed line). The
corresponding values of the Lorentz factor today are $\gamma _{\rm
0}\sim 3.96$, $\gamma _{\rm 0}\sim 17.87$, $\gamma _{\rm 0}\sim 88.56$
and $\gamma _{\rm 0}\sim 176.17$ respectively. The final vacuum
expectation values of the DBI field are $\phi_0/\mpl\sim 1.37$,
$\phi_0/\mpl\sim 0.66$, $\phi_0/\mpl\sim 0.29$ and $\phi_0/\mpl\sim 0.2$
respectively.}  \label{fig:gammattra}
\end{figure}

Let us now study the evolution of the Lorentz factor $\gamma$. It is
represented in Fig.~\ref{fig:gamma} for different initial conditions,
similar to the ones considered in the previous figures. In particular,
the attractor value, valid during the radiation dominated era, $\gamma
_{\rm e}=20$, is clearly seen on this plot. The value $\gamma _{\rm
e}^{\rm cdm}\sim 5.4$, valid during the matter dominated era, can also
be noticed. An interesting point is that the present value of the
Lorentz factor is far from $1$. For the case under consideration, it is
$\gamma _0\sim 3.96$. This means that the non-standard kinetic term
still plays a role even today. As noticed earlier, this can have
important implications for model building issues since a large $\gamma
_{\rm e}$ implies a small vacuum expectation value of the field. We have
studied this point in more details in Fig.~\ref{fig:gammattra} where the
evolution of the Lorentz factor is represented for different values of
$\gamma _{\rm e}$. We notice that the larger $\gamma _{\rm e}$, the
larger the present day value $\gamma _0$ and the smaller $\phi _0$. For
instance, for $\gamma _{\rm e}=1000$, one obtains $\gamma _0\sim 176$
and $\phi _0/\mpl \sim 0.2 <1$. This is certainly a desirable feature
since, usually, vacuum expectation values of the order of the Planck
mass are at the origin of many serious problems as, for instance, a
coupling with the observable sector which violates the constraint on the
presence of a fifth force and/or on the weak equivalence
principle~\cite{Brax:2005uf,Brax:2006dc,Brax:2006kg,Brax:2006np}. On the
other hand, if the vacuum expectation value remains small in comparison
with the Planck mass, then it could be difficult to use the SUGRA
potential model to push the equation of state towards $-1$. Therefore,
we face again the ``no-go theorem'' discussed recently in
Refs.~\cite{Brax:2006dc,Brax:2006kg,Brax:2006np}: what is interesting
from the cosmological point (a large vacuum expectation value in order
to have $\omega _0$ close to $-1$) seems to be incompatible with local
tests of gravity (a large vacuum expectation value usually means a
strong coupling with ordinary matter).

\section{Conclusions}
\label{sec:conclusions}

In this section, we recap our main findings and discuss further issues
that should be investigated. We have studied scenarios where the dark
energy is a scalar field with a DBI kinetic term. We have shown that, if
the brane tension and the potential possess either an exponential or a
power law shape, then there exists scaling solutions that are
attractors. Moreover, if the Lorentz factor is large today, then the
vacuum expectation value of the field can be small in comparison with
the Planck mass. Let us also notice that the fact that the scaling
solutions obtained in this article correspond to brane tensions of the
power-law form is fairly remarkable since this is precisely what happens
in simple string-inspired models. Unfortunately, one needs $\alpha <0$
for which the dark energy density scales faster than that of the
background. Maybe, the most problematic aspects of the scenario is the
fact that the equation of state today is too far from $-1$. 

\par

In order to improve the above described situation, one probably needs
more complicated string inspired models. In particular, one needs shapes
of $T(\phi)$ and $V(\phi)$ that, for $\phi \ll \mpl/\sqrt{\gamma _{\rm
e}}$ are of the power law form (with $\alpha >0$) in order to preserve
the attractor, and, for $\phi \gg \mpl/\sqrt{\gamma _{\rm e}}$, deviate
from this form in order to push the equation of state towards $-1$. Let
us recall at this stage that this is exactly what the SUGRA model does,
the characteristic scale being the Planck mass instead of
$\mpl/\sqrt{\gamma _{\rm e}}$.

\par

The fact that the present-day value of $\gamma _{\rm e}$ can be large is
also an interesting feature of the models under scrutiny. For example,
this implies that the sound velocity squared $c_{\rm s}^2$ can
significantly deviate from $1$ in contrast to the standard case. Indeed,
in the DBI case, the sound velocity squared $c_{\rm s}^2$ is given by the 
following expression
\begin{equation}
\label{eq:soundvel}
c_{\rm s}^2 = \frac{\del p}{\del X}
\left(\frac{\del \rho}{\del X}\right)^{-1}
      = \frac{1}{\gamma_{\rm e}^2}\, ,
\end{equation}
where $p = p(X, \phi)$, $\rho = \rho(X, \phi)$ (and $X =
\dot{\phi}^2/2$). A dark energy component with $c_{\rm s}^2 \ll 1$
implies less power on large scales and, hence, could account for the low
multipoles of the cosmic microwave background anisotropies. Moreover,
this would also produce peculiar features in the matter power spectrum
as discussed in
Refs.~\cite{Erickson:2001bq,DeDeo:2003te,Bean:2003fb}. All these
properties could be used to distinguish the DBI models from the standard
ones. One more general grounds, it is clear that a complete calculation
of the dark energy perturbations could bring new insights on the model. 

\par

Finally, another interesting issue is that of the coupling of dark
energy with the rest of the world. As already mentioned, this is usually
a problem for quintessence because a small mass means a force with a
very long range, see
Refs.~\cite{Brax:2006dc,Brax:2006kg,Brax:2006np}. In some scenarios,
this also implies variation of the constants, as, for instance, the fine
structure constant. However, in the present context, the couplings are
totally different. For example, the coupling with the electromagnetic
field is of the form $\sqrt{\det \left(g_{\mu \nu}+F_{\mu \nu}\right)}$,
where, here, $g_{\mu \nu}$ is the induced metric on the
brane. Therefore, one can maybe expect this issue to be less problematic
than in the standard case. More work is clearly needed in order to draw
definitive conclusions on these matters.

\subsection*{Acknowledgments}

We would like to thank P.~Brax, T.~Chiba, K.~Ichiki, S.~Mukohyama and
J.~Yokoyama for the useful discussions. M.~Y. would like to thank
M.~Lemoine, J.~Martin, and P.~Peter for kind hospitality at Institut
d'Astrophysique de Paris where this work was initiated.  J.~M. would
like to thank the RESCEU (Tokyo University) for warm hospitality. This
work was partially supported by CNRS-JSPS Bilateral Joint Project ``The
Early Universe: a precision laboratory for high energy physics.''
M.~Y. is supported in part by JSPS Grant-in-Aid for Scientific Research
Nos.~18740157 and 19340054.

\bibliography{refdark}

\end{document}